\documentclass[aip,jcp,reprint,amsmath,amssymb]{revtex4-2}
\usepackage{graphicx}
\usepackage{natmove}

\newcommand{\vv}[1]{\mathbf{#1}}

\begin{document}
\title{Approximation of anisotropic pairwise interactions for charged objects using multivariate polynomials and a multipole expansion}

\author{Mohammadreza Fakhraei}
\affiliation{Department of Chemical Engineering, Auburn University, Auburn, AL 36849, USA}

\author{Dylan McElheny}
\affiliation{Department of Chemical Engineering, Auburn University, Auburn, AL 36849, USA}

\author{Chris A. Kieslich}
\email{kieslich@gatech.edu}
\affiliation{Wallace H. Coulter Department of Biomedical Engineering, Georgia Institute of Technology, Atlanta, Georgia 30332, USA}

\author{Michael P. Howard}
\email{mphoward@auburn.edu}
\affiliation{Department of Chemical Engineering, Auburn University, Auburn, AL 36849, USA}

\begin{abstract}
We formulate a physics-informed data-driven method for modeling anisotropic pairwise interactions in the presence of long-ranged electrostatics. The method separates the total interaction into a long-ranged electrostatic interaction that is approximated using a multipole expansion truncated at the dipole level and a short-ranged residual interaction that is approximated using multivariate Chebyshev polynomials fit to measurements from a limited number of configurations. We assess the approach on a sequence of aromatic molecules (benzene, benzonitrile, and phenoxide), finding that it produces satisfactory results using a modest cutoff distance for the short-ranged interaction. This method has applications for modeling complex interactions for, and conducting dynamic simulations of,  synthetic and biological materials with charge.
\end{abstract}

\maketitle

\section{Introduction}
Atomistic molecular dynamics (MD) simulations are a standard tool for understanding how molecular interactions give rise to emergent behavior in materials \cite{frenkel:academic:2002-c1, allen:oxford:2017}. Many important material properties, such as phase behavior \cite{panagiotopoulos:intjthermophys:1994}, dynamics \cite{yeh:jpcb:2004, mysona:physreve:2019-2}, and mesoscopic structures \cite{wang:macro:2001, mysona:physreve:2019}, are sensitive to finite-size effects in simulations and may require large numbers of molecules and long simulation times to be reliably measured. However, large-scale simulations may become impractical or impossible to perform when the molecules of interest are themselves large because of the number of atoms required. In such cases, the simulation can be accelerated using coarse-grained models that reduce the number of degrees of freedom, and hence interaction sites that dominate the computation in the simulation, compared to the atomistic model \cite{noid:annualrevphyschem:2024}. Here, we focus on relatively rigid molecules that can be approximated as a single anisotropic rigid body \cite{nguyen:jcp:2022,allen:molphys:2006,gay:jcp:1981}, coarsening three position coordinates per atom into only three position and three orientation coordinates per body. The interaction between two such bodies can be described by one anisotropic pair potential that is a function of their relative position and orientation, rather than the sum of many isotropic pair potentials between their constituent atoms, potentially leading to a significant reduction in computation. The functional forms of the anisotropic pair potential, however, is not generally known and may depend on the shape and chemistry of the molecules involved.

Data-driven methods have attracted significant interest for numerically approximating anisotropic pairwise interactions due to their mathematical flexibility and generality \cite{nguyen:jcp:2022, campos-villalobos:jcp:2022, wilson:jcp:2023, argun:jcp:2024, hatch:jcp:2024, campos-villalobos:npjcompmat:2024, fakhraei:jpcb:2025, fakhraei:jcp:2026}. These methods have mainly been used to learn the potential energy function representing the interaction from a finite number of samples of a ground-truth model. For example, anisotropic pair interactions for benzene, perylene, and sexithiophene have been approximated using spline functions \cite{nguyen:jcp:2022} and neural networks \cite{wilson:jcp:2023} that were fit to the forces and torques recorded in atomistic MD simulations with the OPLS-AA force field \cite{jorgensen:jacs:1996}. General-purpose machine-learning methods like neural networks have proven to be quite successful but also typically require millions of samples from the ground-truth model, potentially limiting their application when the ground-truth model is expensive to evaluate. To address this data requirement, we recently proposed a method for approximating anisotropic pair potentials using multivariate Chebyshev polynomials \cite{fakhraei:jpcb:2025, fakhraei:jcp:2026}, which have historically been used for data-efficient interpolation of smooth functions \cite{press:cambridge:2007}. We demonstrated that multivariate Chebyshev polynomials accurately approximated anisotropic pair potentials \cite{fakhraei:jpcb:2025}, as well as forces and torques \cite{fakhraei:jcp:2026}, for several shape-anisotropic nanoparticles using significantly fewer samples than previously required for other methods \cite{nguyen:jcp:2022,wilson:jcp:2023,argun:jcp:2024, argun:jcp:2025}.

All of these data-driven methods, including ours, have treated the anisotropic pairwise interaction as a short-ranged potential that can be truncated at a modest cutoff distance, beyond which the interaction is assumed to be negligible. This assumption is reasonable for uncharged, nonpolar molecules, whose interactions are dominated by short-ranged repulsion and dispersion forces that decay rapidly with separation. For charged and/or polar molecules, however, electrostatic interactions decay more slowly and can remain significant at much longer distances. Approximating the anisotropic pairwise interactions for such systems using existing data-driven methods may therefore require either truncating the interaction before it has sufficiently decayed, leading to inaccuracies in the simulation, or increasing the cutoff radius, increasing the computational cost of the simulation.

In this article, we extend our approach based on multivariate Chebyshev polynomial approximations \cite{fakhraei:jpcb:2025, fakhraei:jcp:2026} to anisotropic pairwise interactions in the presence of long-ranged electrostatic interactions. We consider three aromatic molecules (benzene, benzonitrile, and phenoxide) that span a range of electrostatic character, from nonpolar to polar to charged. Leveraging established methods for evaluating long-ranged interactions in MD \cite{ewald,darden:jcp:1993,essmann:jcp:1995,allen:oxford:2017}, we subtract a multipolar expansion of the electrostatic potential energy truncated at the dipole level from the total potential energy evaluated in the ground-truth model, and we approximate only the shorter-ranged residual potential energy with the multivariate Chebyshev polynomials. The total potential energy is then evaluated as the sum of the analytical multipolar expansion of the electrostatic potential energy and the polynomial approximation of the residual potential energy. We find that this physics-informed decomposition can effectively and meaningfully reduce the required cutoff radius for approximating anisotropic pairwise interactions for polar and charged molecules. The approach can be readily extended to coarse-graining using measured forces and torques rather than the potential energy.

The rest of the article is organized as follows. We first describe the atomistic model that served as the ground truth and the construction of the approximate model using multivariate Chebyshev polynomials in Sec.~\ref{sec:methods}. We next present the approximation of the total anisotropic pairwise interaction for the three molecules and identify its limitations, then evaluate the accuracy and practical implications of decomposing the total interaction using a multipolar expansion of the electrostatics in Sec.~\ref{sec:results}. We conclude with a summary of our key results and an outlook in Sec.~\ref{sec:conclusions}.

\section{Model and Methods}
\label{sec:methods}
\subsection{Atomistic Model} 
\begin{figure}[ht]
    \centering
    \includegraphics{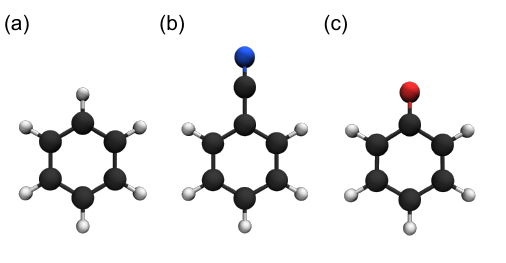}
    \caption{Aromatic molecules in their reference configuration in the $xz$ plane: (a) benzene, (b) benzonitrile, and (c) phenoxide. The meaning of the atom colors are black: carbon, white: hydrogen, blue: nitrogen, and red: oxygen. The images were rendered using VMD 1.9.4 \cite{humphrey:jmolgrp:1996}.}
    \label{fig:molecules}
\end{figure}

We considered three aromatic molecules (Fig.~\ref{fig:molecules}) with similar size, structure, and symmetry but different electrostatic character: benzene, benzonitrile, and phenoxide. All three molecules are a six-member ring but with a different small substituent, so their excluded-volume and dispersion forces were expected to be similar but their electrostatic forces were expected to vary significantly. Specifically, benzene [Fig.~\ref{fig:molecules}(a)] does not have a net charge or permanent dipole moment, so the interaction between two benzene molecules is effectively short-ranged. Benzene therefore served as a baseline of the series with minimal long-ranged electrostatic interactions. Benzonitrile [Fig.~\ref{fig:molecules}(b)] does not have a net charge but it has a dipole moment in the direction of the electron-withdrawing nitrile group, which produces a long-ranged dipolar interaction between two benzonitrile molecules that is absent for benzene. Phenoxide [Fig.~\ref{fig:molecules}(c)] carries a net charge, so it gives rise to an even longer-ranged Coulombic interaction between phenoxide molecules that is absent for both benzene and benzonitrile.

For the purposes of this methodological study, each molecule was treated as a rigid body so that its internal configuration was fixed and its only degrees of freedom were the position of its center of mass and its orientation (rotation) relative to a reference configuration. The reference configuration for each molecule was obtained by minimizing the potential energy calculated using the OPLS-AA force field \cite{jorgensen:jacs:1996}. For benzene and benzonitrile, we built an initial structure for each molecule in Avogadro 1.2.0 \cite{hanwell:jcheminform:2012}, then we used mBuild 0.10.9 \cite{klein:springer:2016} to generate GROMACS (version 2026.0) \cite{abraham:softwarex:2015} coordinate and topology files. For phenoxide, the OPLS-AA force field did not contain the required atom types, so we used the LigParGen server \cite{lipgenweb} with partial charges assigned using the CM1A scheme \cite{jorgensen:pnas:2005, dodda:jpcb:2017} to generate the GROMACS coordinate and topology files directly. Each molecule was then placed in a cubic box, and its energy was minimized using GROMACS to a maximum-force tolerance of $1.0\,{\rm kJ}/({\rm mol}\,{\rm nm})$ using the steepest-descent method and a nonbonded-interaction cutoff distance that was large enough to include all interactions. We then translated and rotated each energy-minimized molecule so that its center of mass was at the origin and the ring lay in the $xz$ plane with an arbitrarily chosen hydrogen for benzene, the nitrile group for benzonitrile, and the oxide group for phenoxide pointed along the $z$ axis. The reference configuration of each molecule is shown in Fig.~\ref{fig:molecules} and included as a coordinate file in the Supplementary Material. Table~\ref{tab:moments} summarizes the calculated net charge and dipole moment about the center of mass for each molecule in its reference configuration, confirming the molecules possess the expected range of electrostatic character.

\begin{table}[ht]
\centering
\caption{Net charge and dipole moment about the center of mass for each molecule in its reference configuration.}
\begin{tabular}{ccc}
molecule & net charge ($e$) &  \ dipole moment (D) \\
\hline
benzene      &  0 & 0.01 \\
benzonitrile &  0 & 3.23 \\
phenoxide    & -1 & 3.73 \\
\end{tabular}
\label{tab:moments}
\end{table}

The intermolecular potential energy for two rigid molecules is the sum of the nonbonded interactions between their constituent atoms. For the OPLS-AA force field \cite{jorgensen:jacs:1996}, the total pairwise potential energy $u = u_{\mathrm{LJ}} + u_{\mathrm{e}}$ comprised a Lennard-Jones excluded-volume and dispersion contribution $u_{\mathrm{LJ}}$ as well as a Coulomb electrostatic contribution $u_{\mathrm{e}}$. Labeling one molecule as 1 and the other as 2, the Lennard-Jones contribution was
\begin{equation}
u_{\mathrm{LJ}} = \sum_{i \in 1} \sum_{j \in 2}
4\varepsilon_{ij}\left[ \left(\frac{\sigma_{ij}}{r_{ij}}\right)^{12}
- \left(\frac{\sigma_{ij}}{r_{ij}}\right)^{6} \right],
\end{equation}
where the indices for the sums run over all atoms $i$ and $j$ in molecules 1 and 2, respectively, $r_{ij} = \lvert \vv{r}_j - \vv{r}_i \rvert$ is the distance between atoms with $\vv{r}_i$ being the position of atom $i$, and $\sigma_{ij}$ and $\varepsilon_{ij}$ are the Lennard-Jones parameters obtained using geometric mixing rules with the per-atom parameters for the force field. The electrostatic contribution was
\begin{equation}
u_{\mathrm{e}} = \sum_{i \in 1} \sum_{j \in 2}
\frac{q_i q_j}{4\pi\varepsilon_0 r_{ij}},
\end{equation}
where $q_i$ is the partial charge of atom $i$ and $\varepsilon_0$ is the permittivity of vacuum. No cutoff was used to evaluate these interactions because the molecules contained only a small number of atoms. From the total energy $u$, the force on atom $i$ is $\vv{f}_{i} = -\partial u / \partial \vv{r}_i$, the net force on molecule 1 is the sum over its atoms,
\begin{equation}
\vv{F}_1 = \sum_{i \in 1} \vv{f}_i,
\end{equation}
and the net force on molecule 2 is $\vv{F}_2 = -\vv{F}_1$. The torque $\boldsymbol{\tau}_1$ on molecule 1 about its center of mass $\vv{R}_1$ is
\begin{equation}
\boldsymbol{\tau}_1 = \sum_{i \in 1} (\vv{r}_i - \vv{R}_1) \times \vv{f}_i,
\end{equation}
and the torque on molecule 2 is $\boldsymbol{\tau}_2 = -\boldsymbol{\tau}_1 + (\vv{R}_2 - \vv{R}_1) \times \vv{F}_1$, where $\vv{R}_2$ is the center of mass for molecule 2. We used OpenMM 8.5.1 \cite{eastman:jpcb:2024} to create pairwise configurations of molecules from their reference configurations and evaluate the associated energy, net forces, and net torques. For this study, we considered only interactions between two molecules of the same type, but the same approach can be applied to mixed pairs of molecules.

\subsection{Approximate Model}
\label{sec:methods:polynomial}
We next constructed an approximate anisotropic pairwise potential energy function $\hat{u}$ as a function of only the relative position and orientation of two molecules using a limited number of samples of the potential energy for the atomistic model $u$. Using the framework we recently developed \cite{fakhraei:jpcb:2025, fakhraei:jcp:2026}, the approximate potential energy was defined as an $N$-term series
\begin{equation}
\hat u(\vv{q}) = \sum_{n=0}^{N-1} c_n \psi_n(\vv{q}),
\label{eq:energy}
\end{equation}
where $\vv{q}$ is the relative position and orientation coordinates of the two molecules, $\psi_n$ is the $n$-th multivariate basis function formed from products of univariate Chebyshev polynomials, and $c_n$ is the coefficient of the $n$-th basis function. The set of coefficients $\{c_n\}$ can be determined in different ways, such as to interpolate the energy or to match the forces and torques in a least-squares sense \cite{fakhraei:jcp:2026}. In this work, we will interpolate the energy at $N$ points where $u$ has been sampled.

We used a transformed and symmetry-aware coordinate system $\vv{q} = (\rho, \theta, \phi, \alpha, \beta, \gamma)$, where $\theta$ and $\phi$ are the azimuthal and polar angles of the position in spherical coordinates, and $\alpha$, $\beta$, and $\gamma$ are body-fixed $z(\alpha)-x(\beta)-z(\gamma)$ Euler angles describing the orientation. The coordinate $\rho$ is a scaled center-to-center distance,
\begin{equation}
\rho = \frac{1/r - 1/r_0}{1/(r_0+r_{\rm c}) - 1/r_0},
\label{eq:rho}
\end{equation}
where $r = |\vv{R}_2 - \vv{R}_1|$ is the distance between the center-to-center distance,
$r_0(\theta, \phi, \alpha, \beta, \gamma)$ is a lower bound on $r$ to exclude heavily overlapping configurations, and $r_{\rm c}$ is a cutoff distance relative to $r_0$. The bounds of each coordinate were $0 \le \rho \le 1$, $0 \le \theta, \gamma \le \pi$, $10^{-5} \le \phi, \beta \le \pi - 10^{-5}$, and $0 \le \alpha \le 2\pi$. The angular coordinates did not span their full typical ranges because the molecules had, to a good approximation, two-fold symmetry about the $z$ axis that allowed configurations outside this domain to be reduced into it using proper rotations and equivalency of Euler angles \cite{nolze:cryst:2015, fakhraei:jpcb:2025}. Additionally, the bounds of $\phi$ and $\beta$ were shifted inward by $10^{-5}$ to avoid points at which expressions for the forces and torques become undefined but have a limit \cite{fakhraei:jcp:2026}. We note that benzene possesses more symmetry than benzonitrile and phenoxide, which would allow us to reduce the angular domain further; however, we intentionally used the same domain for all molecules to facilitate comparison between them.

We defined $r_0$ as the largest distance at which $u_{\rm LJ}$ equaled $5\,k_{\rm B}T$ when $T = 298\,\mathrm{K}$, with $k_{\rm B}$ being the Boltzmann constant. This large positive energy indicates the onset of strong excluded-volume repulsion that makes configurations with $r < r_0$ unlikely to be observed. We based $r_0$ on only $u_{\rm LJ}$, rather than the total potential energy $u$, because it models the excluded volume and so reflects the shape of the molecule. The cutoff distance $r_{\rm c}$ must be chosen so that the interaction being approximated has effectively vanished, and we will discuss how we chose the cutoff for each molecule in Sec.~\ref{sec:results}. We determined $r_0$ numerically  on a dense uniform grid (30 points in $\theta$, 30 points in $\phi$, 60 points in $\alpha$, 30 points in $\beta$, and 30 points in $\gamma$) and evaluated it using multivariate piecewise-linear interpolation. First partial derivatives of $r_0$ were evaluated using central finite differences with step size $10^{-6}$, switching to forward or backward differences at the domain boundaries. 

The set of multivariate-polynomial basis functions $\{\psi_n\}$ was generated as the tensor product of sets of univariate Chebyshev polynomials of the first kind up to a chosen degree for each coordinate. The multivariate sample points at which the approximate potential energy $\hat{u}$ was required to interpolate the true potential energy $u$ were taken to be the tensor product of the extrema of the highest-degree univariate Chebyshev polynomial for each coordinate, so there were hence also $N$ multivariate sample points. To keep consistent with our motivation of using limited sampling, we capped the total number of multivariate sample points to be $N \le 10^5$. We started by choosing the number of basis functions to use for the scaled distance coordinate $\rho$, which in our experience is an important coordinate to faithfully capture, and we found 9 univariate basis functions to be sufficient for all molecules in preliminary tests. We then estimated the difficulty of approximating the energy along each angular coordinate. We generated $10^4$ configurations at random within the approximation domain. For each configuration, we took each angular coordinate in turn and constructed a univariate interpolant with 7 basis functions while all other coordinates were held fixed, and we calculated the root mean squared error (RMSE) at the value of the angle being interpolated across all configurations (Table S1). Because $r$ was held fixed, some required sample points had $r < r_0$, so we capped the energy at $u(r_0)$ for these points. The RMSE serves as a proxy for the resolution each coordinate requires. We found the error was consistently largest along $\phi$, comparable and intermediate along $\theta$ and $\beta$, and smallest along $\gamma$, so we chose the number of univariate basis functions to be 7 for $\phi$, 6 for $\theta$ and $\beta$, and 5 for $\gamma$. We last set the number of univariate basis functions for $\alpha$ to be 8, giving a total of $N = 90,720$ multivariate basis functions and sample points. The set of coefficients $\{c_n\}$ were determined by solving the linear system of equations for interpolation.

\section{Results and Discussion}
\label{sec:results}
\begin{figure*}[ht]
    \includegraphics{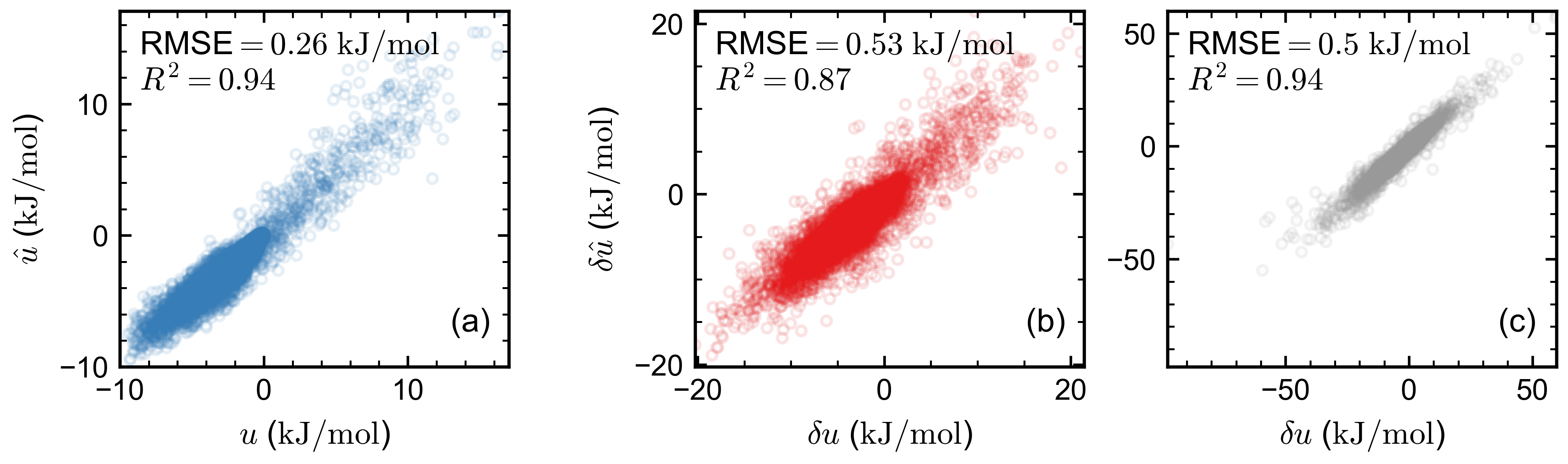}
    \caption{Parity analysis of the approximated (a) total potential energy $\hat{u}$ for benzene, (b) residual potential energy $\delta\hat{u}$ for benzonitrile, and (c) residual potential energy $\delta\hat{u}$ for phenoxide compared to their respective values, $u$ and $\delta u$, for the atomistic model.}
    \label{fig:energy_parity}
\end{figure*}

\begin{figure*}[ht]
    \includegraphics{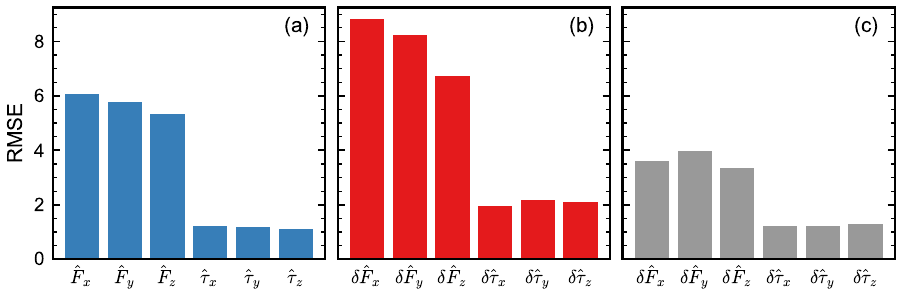}
    \caption{RMSE of approximated forces and torques for (a) benzene, (b) benzonitrile, and (c) phenoxide. The unit of RMSE is ${\rm kJ}/({\rm mol}\,{\rm nm})$ for forces and ${\rm kJ}/{\rm mol}$ for torques. The corresponding parity analysis is shown in Figs.~S2, S4, and S9.}
    \label{fig:rmse_bar}
\end{figure*}

We first assessed the effectiveness of our framework for approximating pairwise interactions for benzene [Fig.~\ref{fig:molecules}(a)], the molecule with the shortest-ranged interactions that served as a baseline for this work. We used a cutoff distance of $r_{\rm c}=1\,{\rm nm}$ for benzene based on the decay of the potential energy with separation distance $r-r_0$ for three different configurations [Fig.~S1(a)--(c)]. This cutoff distance is also reasonable based on the Lennard-Jones parameters for benzene. We then sampled the potential energy $u$ for the atomistic model at the points prescribed by the chosen Chebyshev polynomials (Sec.~\ref{sec:methods:polynomial}) and interpolated these values to obtain the coefficients in eq.~\eqref{eq:energy}. To assess the accuracy of $\hat{u}$, we generated an additional $10^5$ test configurations at random within the approximation domain, and we calculated the energy for both the atomistic model and the approximate model at these points. We found that the approximate potential energy $\hat{u}$ reproduced the true potential energy $u$ for the test configurations with high accuracy, as indicated by a parity analysis [Fig.~\ref{fig:energy_parity}(a)] with an RMSE less than $1\%$ of the range of the true values and a Pearson correlation coefficient $R^2$ of 0.94. We also computed the true and approximate forces and torques for the test configurations, finding again small RMSE [Fig.~\ref{fig:rmse_bar}(a)] compared to the range of observed values and good parity (Fig.~S2).

We then turned our attention to benzonitrile [Fig.~\ref{fig:molecules}(b)], whose nitrile group withdraws electron density and gives it a permanent dipole moment of $3.23\,\mathrm{D}$. This dipole moment has a direct consequence for a data-driven approximation: benzonitrile's leading far-field electrostatic term is the interaction between two permanent dipoles, which decays as $1/r^3$ and so is significant to much larger distances than $u_{\rm LJ}$, which decays as $1/r^6$ for pairs of atoms. A longer cutoff distance $r_{\rm c}$ would therefore be needed for benzonitrile, and we identified $r_{\rm c} = 2\,{\rm nm}$ to be suitable based on potential energy traces with respect to separation distance for three test configurations [Figs.~S1(d)--(f)].

Increasing the cutoff distance from 1 nm to 2 nm for benzonitrile has an important practical implication for using the approximate potential energy $\hat{u}$ in a simulation. Evaluating eq.~\eqref{eq:energy} involves a significant amount of calculation because of the required coordinate transformations and the number of terms, so the number of pairwise interactions evaluated $N_{\rm p}$ is an important factor in simulation speed. To estimate the impact of changing the cutoff, we assume a uniform distribution of relative positions and orientations for $r_0 \lesssim r \le r_0 + r_{\rm c}$, so $N_{\rm p}$ is proportional to the volume of the configuration space,
\begin{equation}
N_{\rm p} \propto \int \big[(r_0 + r_{\rm c})^3 - r_0^3\big]\,
\sin\phi\,\sin\beta\; {\rm d}\theta\,{\rm d}\phi\,{\rm d}\alpha\,{\rm d}\beta\,{\rm d}\gamma,
\end{equation}
where the limits of integration are $0\le \theta,\alpha,\gamma < 2\pi$ and $0 \le \phi, \beta \le \pi$. We evaluated this integral numerically for benzonitrile within the domain of the angles for $\hat{u}$ using the trapezoid rule on the uniform grid of values of $r_0$, as the full integral is essentially proportional by symmetry. We found that $N_{\rm p}$ was approximately 4.6 times larger when $r_{\rm c} = 2\,{\rm nm}$ than when $r_{\rm c}=1\,{\rm nm}$, a substantial increase. We further note that for charged molecules such as phenoxide [Fig.~\ref{fig:molecules}(c)], the required cutoff distance becomes wholly impractical (on the order of tens of nanometers [Fig.~S1(g--i)]), as for standard electrostatic interactions in atomistic MD.

These observations motivated us to implement a physics-informed strategy for handling long-ranged electrostatics in data-driven models for anisotropic pairwise interactions without expanding $r_{\rm c}$ significantly. Note that only the electrostatic contribution to the potential energy is long-ranged, and $u_{\rm e}$ can be approximated at large distances using a truncated multipolar expansion \cite{jackson:wiley:1999, allen:oxford:2017},
\begin{equation}
u_{\rm mp} \approx u_{\rm mm} + u_{\rm md} + u_{\rm dd},
\end{equation}
where each term depends on the net charge (or monopole) $Q_m$ and/or the dipole $\boldsymbol{\mu}_m$ of molecule $m$ \cite{allen:oxford:2017},
\begin{align}
Q_m &= \sum_{i \in m} q_i \\
\boldsymbol{\mu}_m &= \sum_{i \in m} q_i (\vv{r}_i - \vv{R}_m).
\end{align}
The three terms in $u_{\rm mp}$ are the monopole--monopole interaction,
\begin{equation}
u_{\rm mm} = \frac{Q_1 Q_2}{4\pi\varepsilon_0 r},
\end{equation}
decaying as $1/r$; the monopole--dipole interaction,
\begin{equation}
u_{\rm md} = \frac{1}{4\pi\varepsilon_0 r^2}
\big[ Q_2\,(\boldsymbol{\mu}_1 \cdot \hat{\vv{r}}) - Q_1\,(\boldsymbol{\mu}_2 \cdot \hat{\vv{r}}) \big],
\end{equation}
decaying as $1/r^2$ with $\hat{\vv{r}} = (\vv{R}_2 - \vv{R}_1)/r$; and the dipole--dipole interaction,
\begin{equation}
u_{\rm dd} = \frac{1}{4\pi\varepsilon_0 r^3}
\big[ \boldsymbol{\mu}_1 \cdot \boldsymbol{\mu}_2
- 3\,(\boldsymbol{\mu}_1 \cdot \hat{\vv{r}})(\boldsymbol{\mu}_2 \cdot \hat{\vv{r}}) \big],
\end{equation}
decaying as $1/r^3$. The multipolar expansion can be carried to a higher order, for example by including the quadrupole, at the cost of adding more terms that become increasingly complex to evaluate \cite{aguado:2003}. We consider only up to the dipole here because it is the minimum needed to obtain both forces and torques.

We proposed to approximate only the residual potential energy
\begin{equation}
\delta u = u - u_{\rm mp},
\end{equation}
using a data-driven model $\delta\hat{u}$ that interpolates $\delta u$ rather than the total potential energy. Because $u_{\rm mp}$ should approximate the electrostatic interactions well at large distances, the residual $\delta u$ should decay more rapidly with $r$ than $u$ itself, allowing a shorter cutoff distance $r_{\rm c}$ to be used for $\delta \hat{u}$ than for $\hat{u}$. Further, even though $u_{\rm mp}$ becomes a less accurate model of the electrostatic interactions at short distances, $\delta\hat{u}$ can effectively serve as a corrector that approximates both $u_{\rm LJ}$ and the difference $u_{\rm e} - u_{\rm mp}$. Hence, the data-driven approximation $\delta\hat{u}$ is used for the short-ranged interactions, and a physics-based approximation $u_{\rm mp}$ is used for the long-ranged interactions. The full interaction is recovered by adding the two, $\hat u = \delta \hat{u} + u_{\rm mp}$.

We applied this strategy to benzonitrile, whose multipolar expansion contained only the dipole--dipole interaction $u_{\rm dd}$ because it has no net charge. As anticipated, the residual potential energy $\delta u$ decayed faster as a function of separation distance [Figs.~S3(a)--(c)] than $u$ [Figs.~S1(d)--(f)] for the same three configurations. The residual potential energy reached a negligible value at a separation distance of about 1 nm, half the distance required for the total potential energy and comparable to that used for benzene. We therefore set $r_{\rm c} = 1\,\mathrm{nm}$ and constructed an approximation of the residual potential energy $\delta \hat{u}$ from samples of $\delta u$ following the same procedure used for $\hat{u}$. When evaluated on the test configurations, $\delta \hat{u}$ approximated the residual interaction as shown in a parity analysis of the residual potential energy [Fig.~\ref{fig:energy_parity}(b)] and the corresponding residual forces and torques [Figs.~\ref{fig:rmse_bar}(b) and Fig.~S4]. The absolute RMSE was larger and $R^2$ was somewhat smaller for benzonitrile than for benzene, but the RMSE relative to the range of true values was comparable. The total potential energy, forces, and torques were also approximated with high accuracy (Figs.~S5 and S6), indicating that the long-range electrostatics were captured accurately by $u_{\rm mp}$.

\begin{figure}[ht]
    \includegraphics{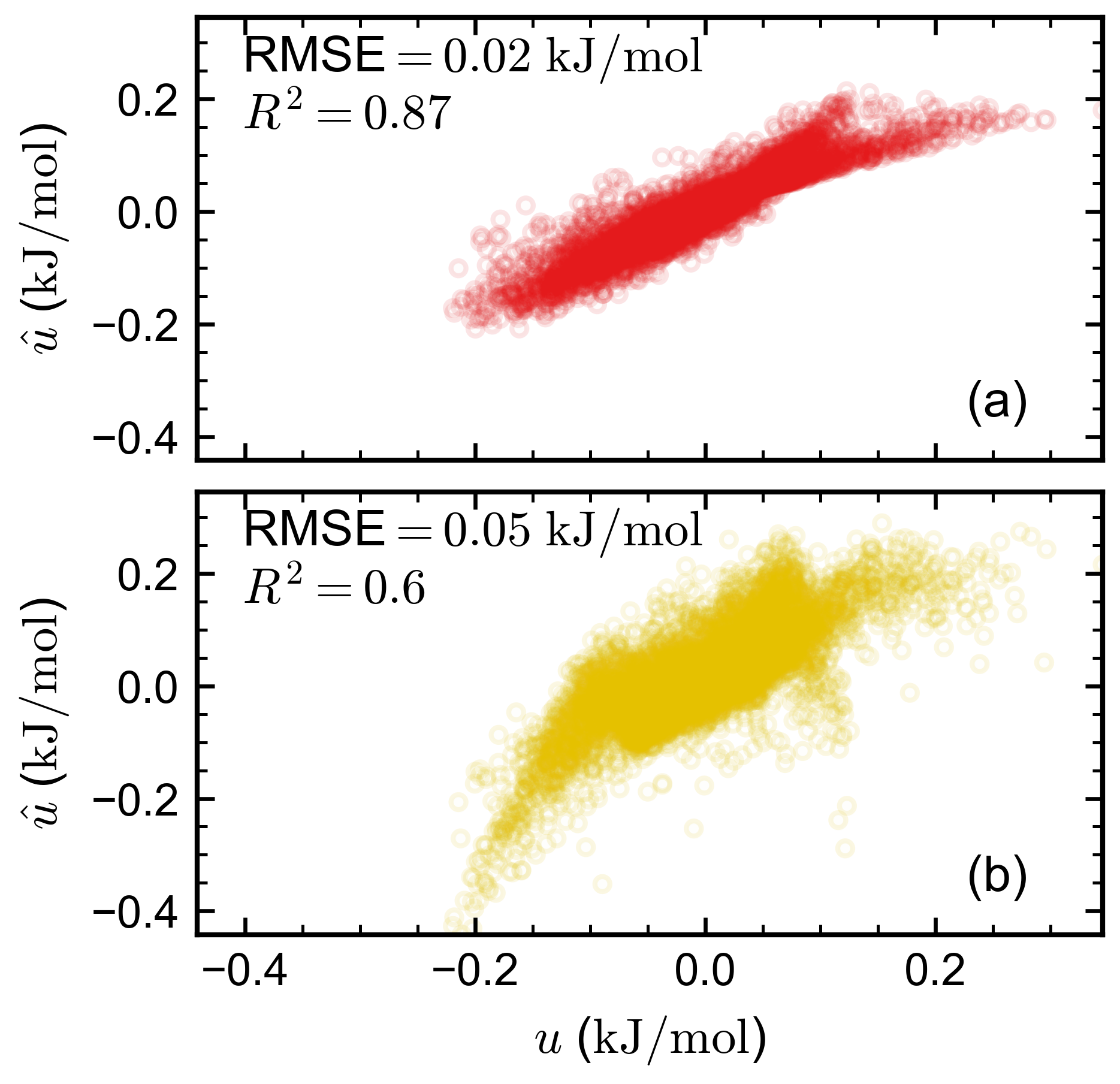}
    \caption{Parity analysis of the approximated total potential energy $\hat{u}$ for benzonitrile when $1\,{\rm nm} < r - r_0 \le 2\,{\rm nm}$ using (a) multipolar expansion and (b) data-driven model.}
    \label{fig:long_range}
\end{figure}

We compared our proposed strategy of approximating $\delta\hat{u}$ with a direct approximation $\hat{u}$ of the total potential energy using the larger cutoff distance, $r_{\rm c} = 2\,\mathrm{nm}$. The approximation $\hat{u}$ showed comparable accuracy for test configurations within the cutoff distance for $\delta \hat{u}$, $r_{\rm c} = 1\,{\rm nm}$ (Fig.~S7). This result can be understood from the distribution of the sample points under the scaled distance coordinate [eq.~\eqref{eq:rho}]. Because $\rho$ is linear in $1/r$, sample points are placed densely at small separations and more sparsely at larger separations. Increasing the cutoff from 1 nm to 2 nm redistributes the same number of sample points over a larger distance; however, the near-contact region, where the interaction varies most rapidly and is most difficult to approximate, remains densely sampled in both cases. The sample points at large separation move farther apart, but the interactions are weaker and smoothly varying in this region and so easier to approximate. To understand the accuracy of $\hat{u}$ at larger separations, we generated an additional $10^4$ test configurations restricted to $1\,\mathrm{nm} < r - r_0 \le 2\,\mathrm{nm}$. For our proposed strategy, only $\delta\hat{u} = 0$ in this region and only $u_{\rm mp}$ is nonzero. This test hence compares the accuracy of the physics-informed approximation $u_{\rm mp}$ with the data-driven model $\hat{u}$. We found that the multipolar expansion was substantially more accurate than the data-driven model (Fig.~\ref{fig:long_range}). This analysis shows that approximating the long-range electrostatic interactions analytically not only shortens the required cutoff distance for the data-driven model but also more faithfully models them than simply extending the distance covered. 

We could also, in principle, approximate the total potential energy within the shorter cutoff distance, $r_{\rm c} = 1\,\mathrm{nm}$. For completeness, we constructed such a model and assessed its accuracy, finding that it also had good accuracy comparable to interpolating $\delta u$ (Fig.~S8). However, we emphasize that $u$ and its derivatives are not zero at this cutoff distance, leading to discontinuities in forces and torques that are problematic for MD simulations. These discontinuities might be addressed by adding a switching or smoothing function \cite{}, but this function modifies the interaction and also may not be reasonable to use if the discontinuities are large. Approximating the residual $\delta u$ avoids this problem because $\delta u$ is assumed to decay to zero within the cutoff distance.

We last applied our strategy to phenoxide, whose multipolar expansion has the monopole--monopole term $u_{\rm mm}$ as its leading term, which decays as $r^{-1}$, meaning that the total potential energy cannot be truncated within any practical cutoff distance [Fig.~S1(g)--(i)]. Subtracting $u_{\rm mp}$ reduces the cutoff distance required for $\delta \hat{u}$ to 2 nm based on traces of $\delta u$ with respect to separation distance for three configurations [Fig.~S3(d)--(f)]. This cutoff distances is larger than for benzonitrile and might be shortened by continuing the expansion to include quadrupole interactions, as the monopole--quadrupole term also decays as $1/r^3$, but the computational cost of evaluating $u_{\rm mp}$ would increase significantly. We accordingly used $r_{\rm c} = 2\,\mathrm{nm}$ to approximate $\delta \hat{u}$ for phenoxide, and we found in a parity analysis [Fig.~\ref{fig:energy_parity}(c)] that $\delta\hat{u}$ for phenoxide had a comparable absolute RMSE but smaller relative RMSE, as well as a larger $R^2$, than for benzonitrile. The residual forces and torques were also of comparable accuracy to the approximations for the other two molecules [Fig.~\ref{fig:rmse_bar}(c)].

\section{Conclusions}
\label{sec:conclusions}
We have extended our data-driven method for approximating anisotropic pairwise interactions using limited sampling \cite{fakhraei:jpcb:2025, fakhraei:jcp:2026} to objects with electrostatic interactions, such as charged or dipolar molecules. Directly approximating the total interaction with existing data-driven methods would require a large or impractical cutoff radius. We addressed this challenge by decomposing the interaction into a long-ranged part, which we approximated by a multipolar expansion of the electrostatics truncated at the dipole level, and a short-ranged part that captures the residual interaction, which we approximated using multivariate Chebyshev polynomials. We assessed this strategy using three aromatic molecules (benzene, benzonitrile, and phenoxide) with a similar chemical structure but different electrostatic character, finding that it achieved good accuracy at both short and long distances. The same strategy could be applied to more complex molecules whose interactions are dominated by long-ranged electrostatics, such as peptides and proteins with charged and polar residues, charged colloidal particles, and ionic surfactants. 

We approximated the potential energy between two molecules calculated using a ground-state atomistic model in this study, but a similar approach could be used to perform force and torque matching for coarse-graining applications \cite{nguyen:jcp:2022, wilson:jcp:2023}. In this context, some potential limitations are use of the ground-state structure and the assumption of rigidity as well as  the application of symmetry to this structure because real molecules have rotational and vibrational fluctuations that break these assumptions. The severity of these limitations are expected to depend on the molecular structure and the thermophysical properties of interest, so MD simulations of the approximate model are hence needed to assess them. An implementation of eq.~\eqref{eq:energy} for MD simulations on high-performance computing resources is needed to do so. We note that potentials with related functional forms but different applications, such as the ChIMES force fields \cite{lindsey:jchemtheorycomp:2017, lindsey:jchemtheorycomp:2019, lindsey:jchemphys:2020, lindsey:jchemphys-2:2020, pham:jchemphys:2020, lindsey:jchemphys:2023,goldman:jchemphys:2023,lindsey:natmat:2025}, have been used successfully deployed at scale, and we are currently developing an implementation of our potential. Further, there are well-established methods for evaluating the terms in the multipole expansion of the electrostatics up to the dipole level using Ewald summation techniques \cite{ewald, darden:jcp:1993, essmann:jcp:1995, allen:oxford:2017} that are available in widely-used simulation software \cite{LAMMPS, eastman:jpcb:2024}. Hence, we anticipate it will be possible to carry out such simulations in the near future.

\section*{Supplementary Material}
See the supplementary material for the analysis used to determine the cutoff distances and select the univariate polynomial degree for each coordinate, as well as parity analysis of the energy, forces, and torques for all three molecules.

\section*{Conflicts of interest}
The authors have no conflicts to disclose.

\section*{Data Availability}
The data that support the findings of this study are available from the authors upon reasonable request. The code for constructing and fitting the multivariate polynomials is available at \url{https://github.com/mphowardlab/smolyay}.

\section*{Acknowledgments}
This material is based upon work supported by the National Science Foundation under Award Nos.~2223084 (D.M.) and 2310724 (M.F.~and M.P.H.). We acknowledge support from the National Institutes of Health under Award No.~R35GM147164 (C.A.K.). This work was completed with resources provided by the Auburn University Easley Cluster.

\bibliography{references}

\end{document}


\title{Supplementary material for ``Approximation of anisotropic pairwise interactions for charged objects using multivariate polynomials and a multipole expansion''}

\author{Mohammadreza Fakhraei}
\affiliation{Department of Chemical Engineering, Auburn University, Auburn, AL 36849, USA}

\author{Dylan McElheny}
\affiliation{Department of Chemical Engineering, Auburn University, Auburn, AL 36849, USA}

\author{Chris A. Kieslich}
\email{kieslich@gatech.edu}
\affiliation{Wallace H. Coulter Department of Biomedical Engineering, Georgia Institute of Technology, Atlanta, Georgia 30332, USA}

\author{Michael P. Howard}
\email{mphoward@auburn.edu}
\affiliation{Department of Chemical Engineering, Auburn University, Auburn, AL 36849, USA}

\maketitle
\begin{table}[!h]
    \centering
    \caption{RMSE (in kJ/mol) for one-dimensional interpolation of energy for each angular coordinate using 7 points with other coordinates held fixed at $10^4$ random configurations.}
    \begin{tabular*}{3in}{@{\extracolsep{\fill}}lcccc@{}}
    \centering
    molecule     & $\theta$ & $\phi$ & $\beta$ & $\gamma$\\
    \hline
    benzene      & 0.65 & 0.70 & 0.44 & 0.36\\
    benzonitrile & 0.73 & 2.40 & 0.88 & 0.44\\
    phenoxide    & 0.26 & 0.48 & 0.20 & 0.18\\
    \end{tabular*}
\end{table}

\begin{figure}[!h]
    \includegraphics{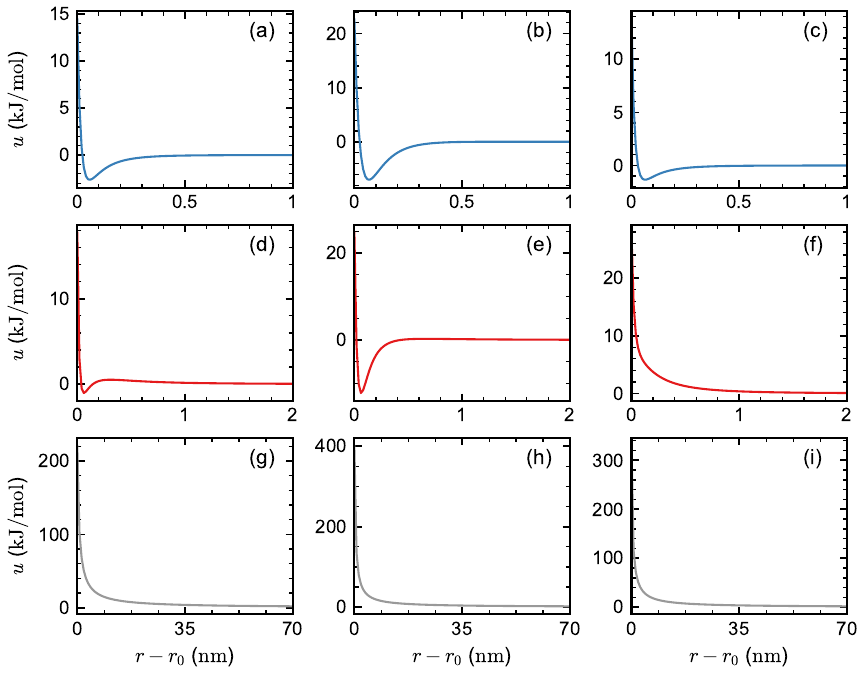}
    \caption{The total energy $u$ as a function of separation distance $r-r_0$ for three configurations of (a--c) benzene, (d--f) benzonitrile, and (g--i) phenoxide. The relative position and orientation coordinates ($\theta,\phi,\alpha,\beta,\gamma$) for (a,d,g) are  $(0,\pi/2,0,0,0)$, for (b,e,h) are $(\pi/2,\pi/2,0,0,0)$, and for (c,f,i) are $(0,0,0,\pi,0)$.}
    \label{fig:u_cutoff}
\end{figure}

\begin{figure}[!h]
    \includegraphics[width=\linewidth]{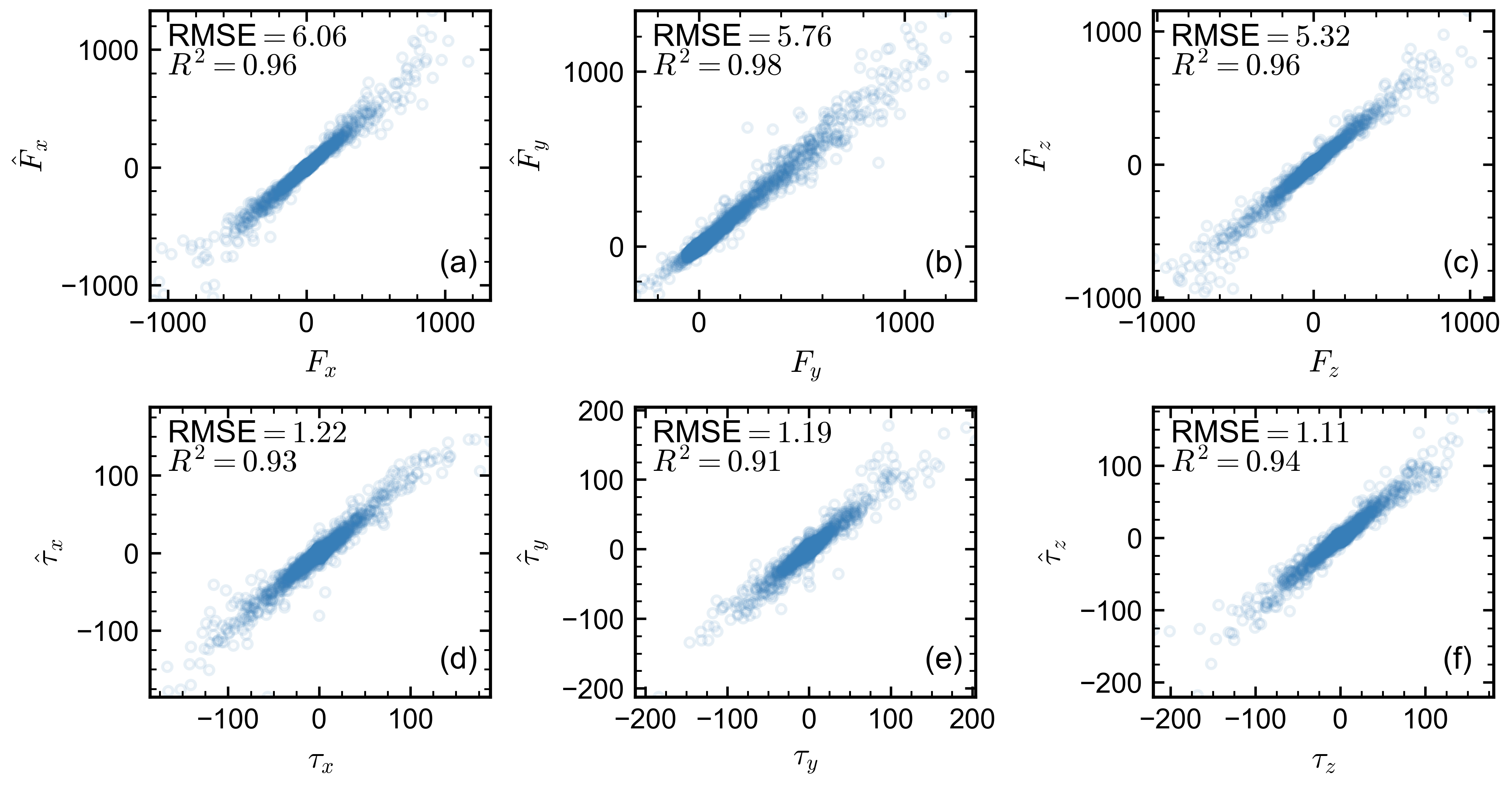}
    \caption{Parity analysis of the approximated (a) $x$-component of the force vector $\hat{F}_x$, (b) $y$-component of the force vector $\hat{F}_y$, (c) $z$-component of the force vector $\hat{F}_z$, (d) $x$-component of the torque vector $\hat{\tau}_x$, (e) $y$-component of the torque vector $\hat{\tau}_y$, and (f) $z$-component of the force vector $\hat{\tau}_z$ for benzene compared to their respective values, $F_x$, $F_y$, $F_z$, $\tau_x$, $\tau_y$, and $\tau_z$ for the atomistic model. The unit of force is ${\rm kJ}/({\rm mol}\,{\rm nm})$, and the unit of torque is ${\rm kJ}/{\rm mol}$.}
\end{figure}

\begin{figure}[!h]
    \includegraphics{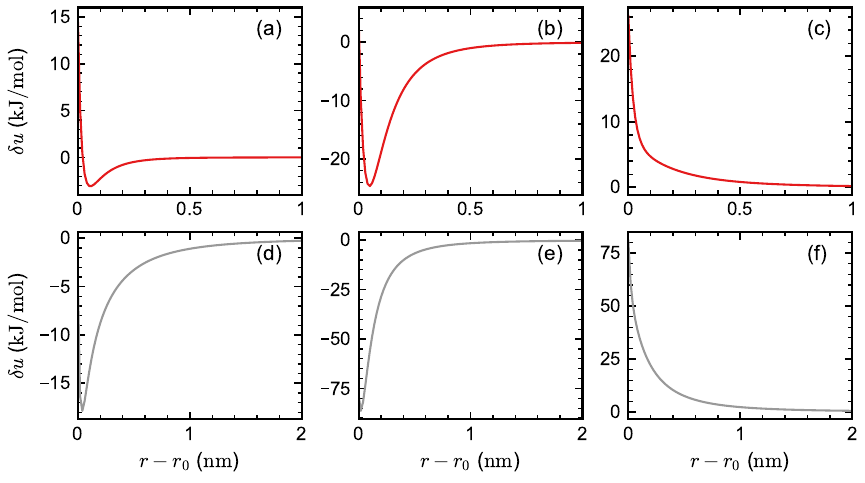}
    \caption{The residual energy $\delta u$ as a function of separation distance $r-r_0$ for three configurations of (a--c) benzonitrile and (d--f) phenoxide. The relative position and orientation coordinates are the same for each column as in Fig.~\ref{fig:u_cutoff}.}
\end{figure}

\begin{figure}[!h]
        \includegraphics[width=\linewidth]{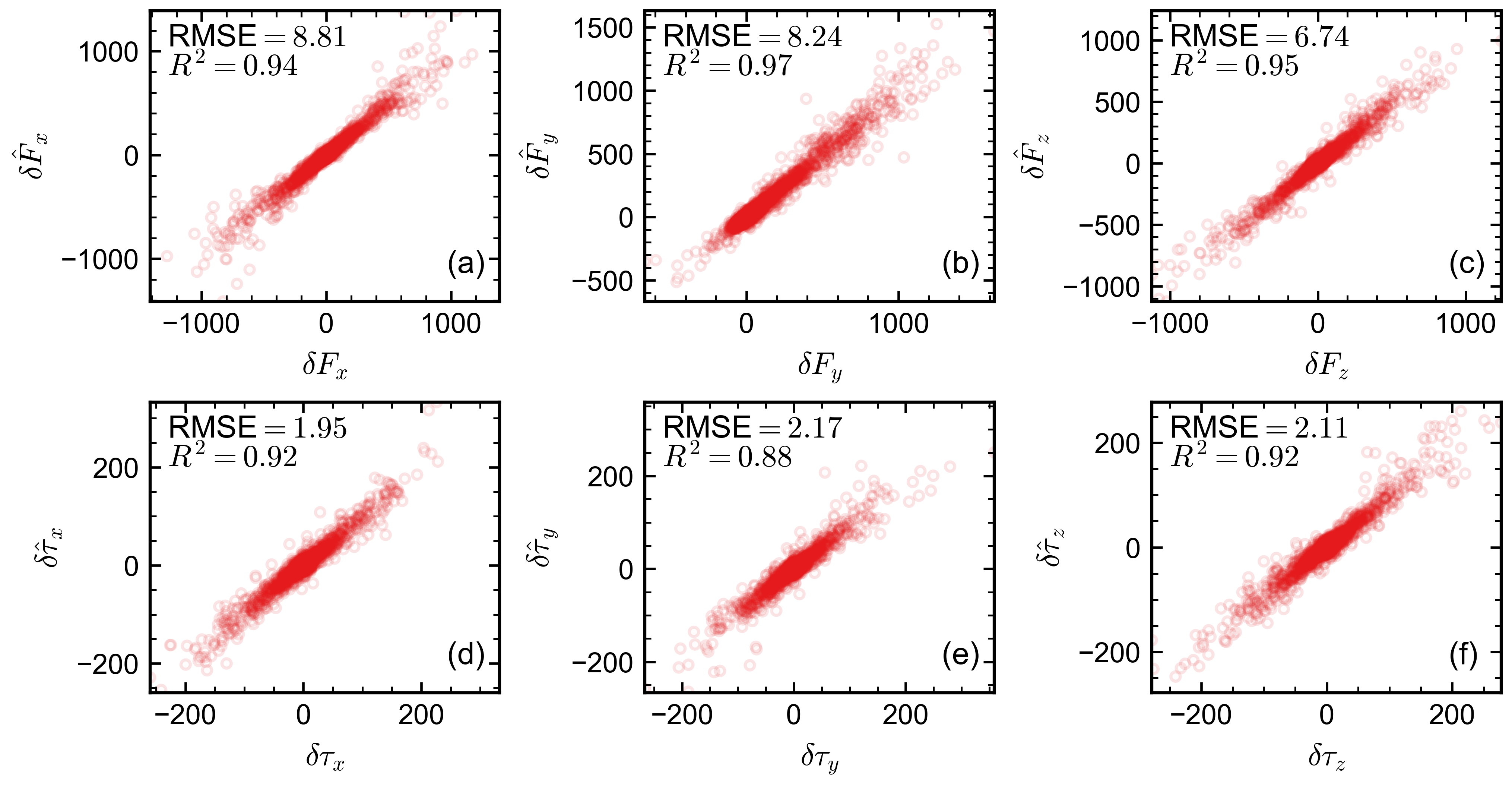}
        \caption{Parity analysis of the approximated (a) $x$-component of the residual force vector $\delta \hat{F}_x$, (b) $y$-component of the force vector $\delta \hat{F}_y$, (c) $z$-component of the force vector $\delta \hat{F}_z$, (d) $x$-component of the torque vector $\delta \hat{\tau}_x$, (e) $y$-component of the torque vector $\delta \hat{\tau}_y$, and (f) $z$-component of the force vector $\delta \hat{\tau}_z$ for benzene compared to their respective values, $\delta F_x$, $\delta F_y$, $\delta F_z$, $\delta \tau_x$, $\delta \tau_y$, and $\delta \tau_z$ for the atomistic model. The approximation is trained by interpolation of residual interaction energy $\delta u$. The unit of forces is ${\rm kJ}/({\rm mol}\,{\rm nm})$, and the unit of torque is ${\rm kJ}/{\rm mol}$.}
        \label{fig:ft_parity_benzo}
\end{figure}

\begin{figure}[!h]
        \includegraphics{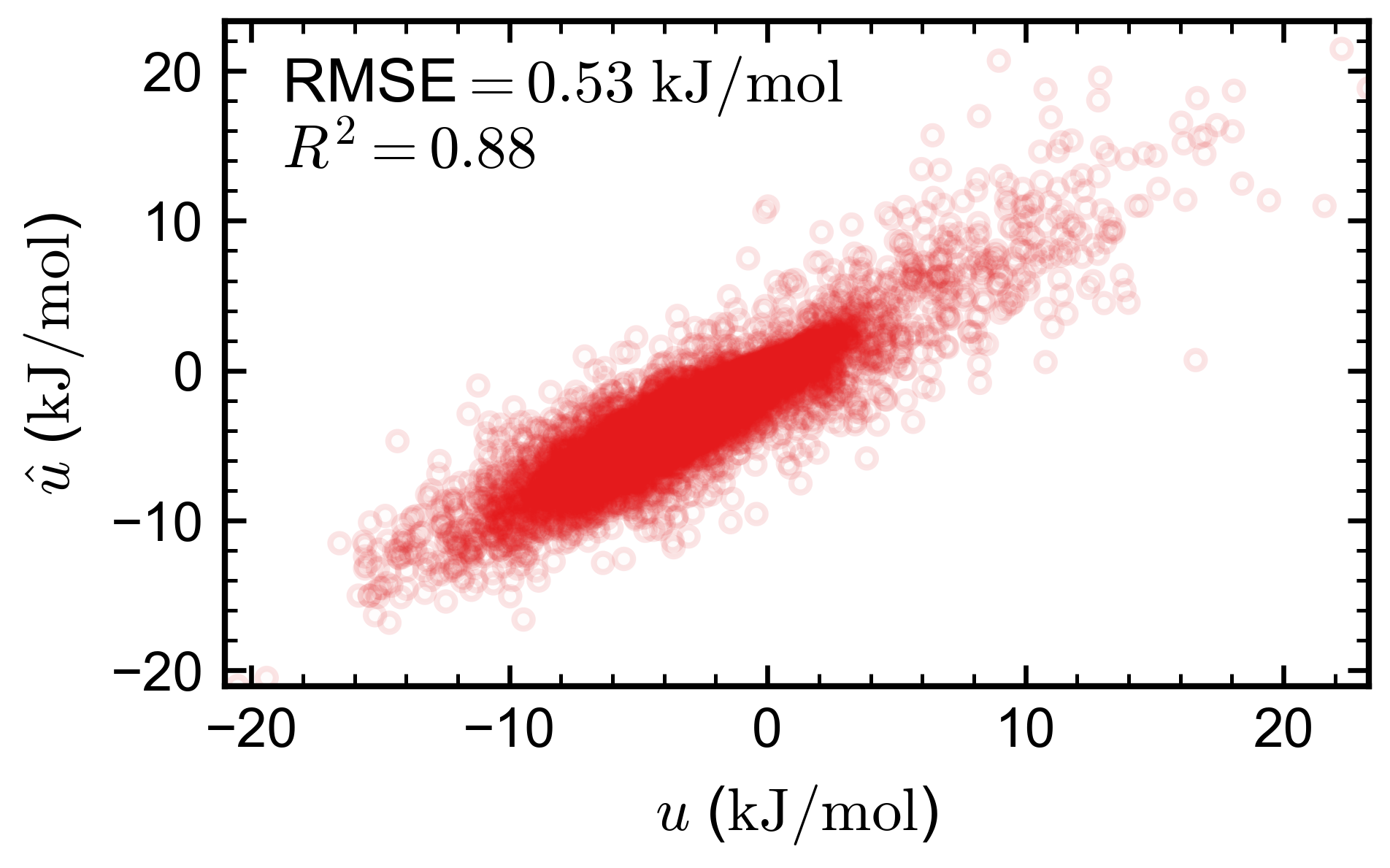}
        \caption{Parity analysis of the approximated total potential energy for benzonitrile $\hat{u}$ compared to its value, $u$, for the atomistic model. The approximation was constructed by interpolation of the residual potential energy $\delta u$, \textit{i.e.}, $\hat{u} = \delta \hat{u} + u_{\rm mp}$ .}
        \label{fig:ut_parity_benzo}
\end{figure}

\begin{figure}[!h]
        \includegraphics[width=\linewidth]{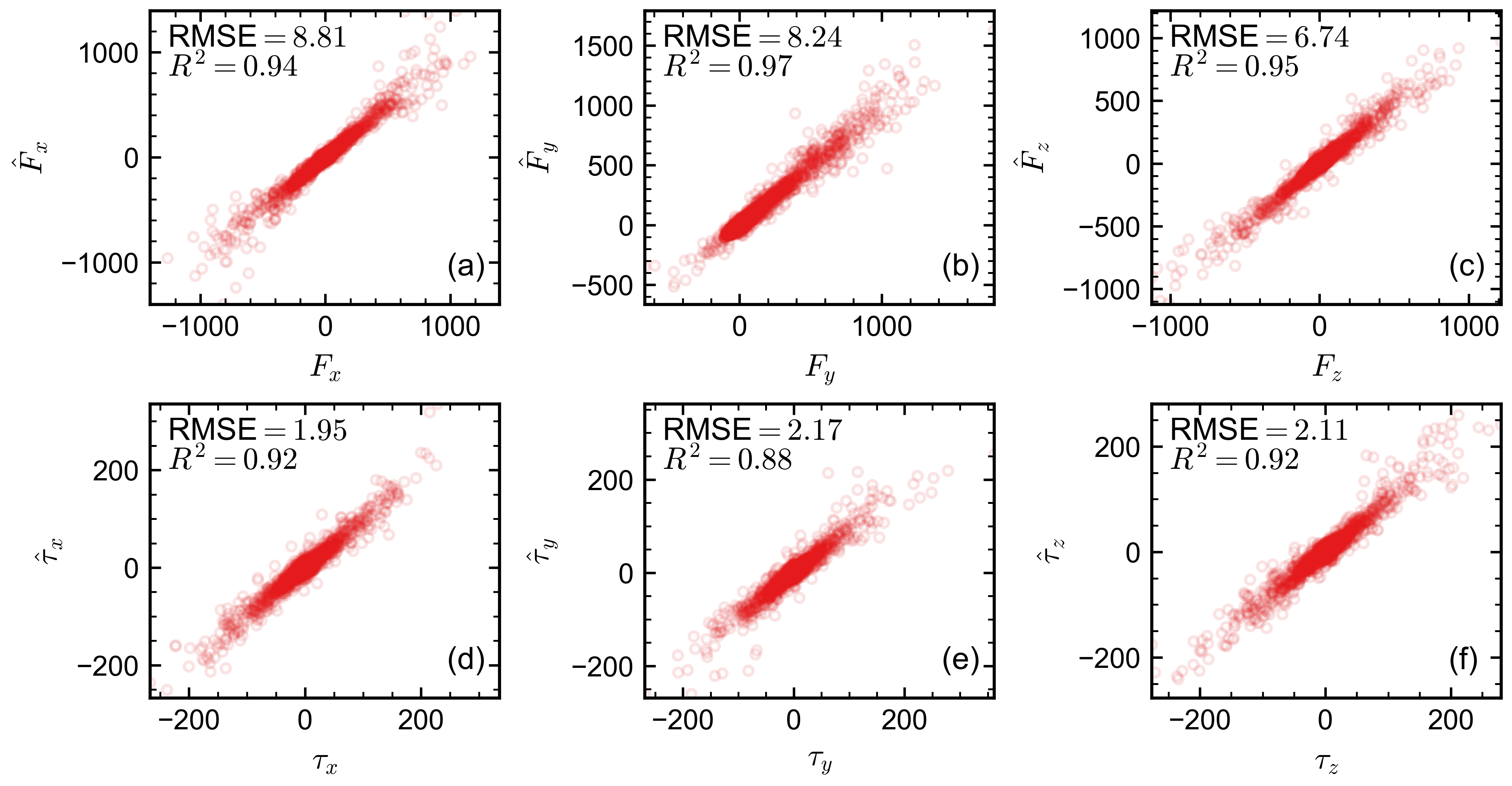}
        \caption{Parity analysis of the approximated (a) $x$-component of the force vector $\hat{F}_x$, (b) $y$-component of the force vector $\hat{F}_y$, (c) $z$-component of the force vector $\hat{F}_z$, (d) $x$-component of the torque vector $\hat{\tau}_x$, (e) $y$-component of the torque vector $\hat{\tau}_y$, and (f) $z$-component of the force vector $\hat{\tau}_z$ for benzonitrile compared to their respective values, $F_x$, $F_y$, $F_z$, $\tau_x$, $\tau_y$, and $\tau_z$ for the atomistic model. The approximation is trained by interpolation of residual interaction energy $\delta u$. The unit of force is ${\rm kJ}/({\rm mol}\,{\rm nm})$, and the unit of torque is ${\rm kJ}/{\rm mol}$.}
        \label{fig:ft_t_parity_benzo}
\end{figure}

\begin{figure}[!h]
        \includegraphics{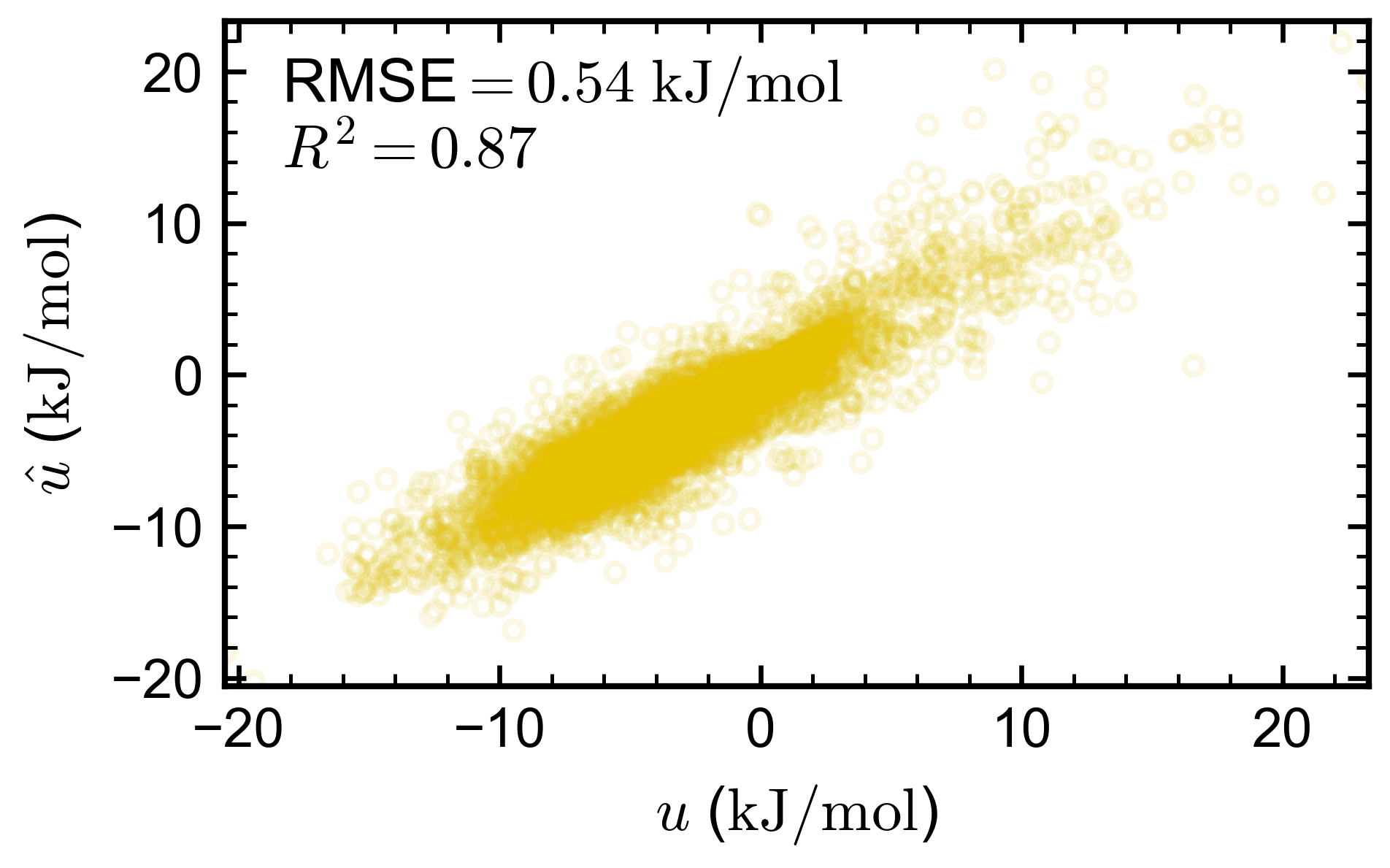}
        \caption{Parity analysis of the approximated total potential energy $\hat{u}$ compared to its value $u$ for the atomistic model of benzonitrile. The approximation was trained within $r_{\rm c} = 2\,{\rm nm}$ and by interpolation of total potential energy $u$.}
\end{figure}

\begin{figure}[!h]
        \includegraphics{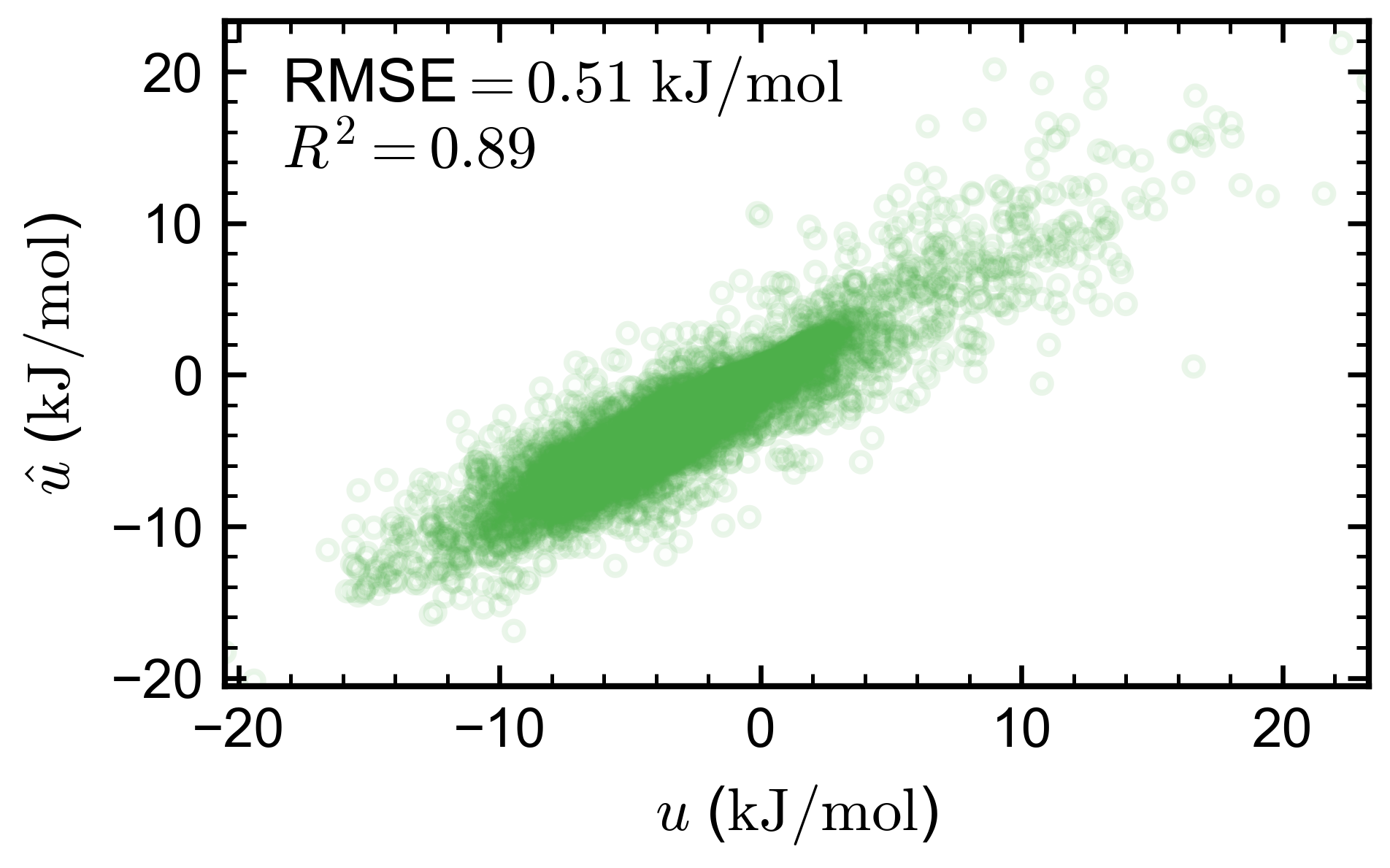}
        \caption{Parity analysis of the approximated total potential energy $\hat{u}$ compared to its value $u$ for the atomistic model of benzonitrile. The approximation was trained within $r_{\rm c} = 1\,{\rm nm}$ and by interpolation of total potential energy $u$.}
\end{figure}

\begin{figure}[!h]
        \includegraphics[width=\linewidth]{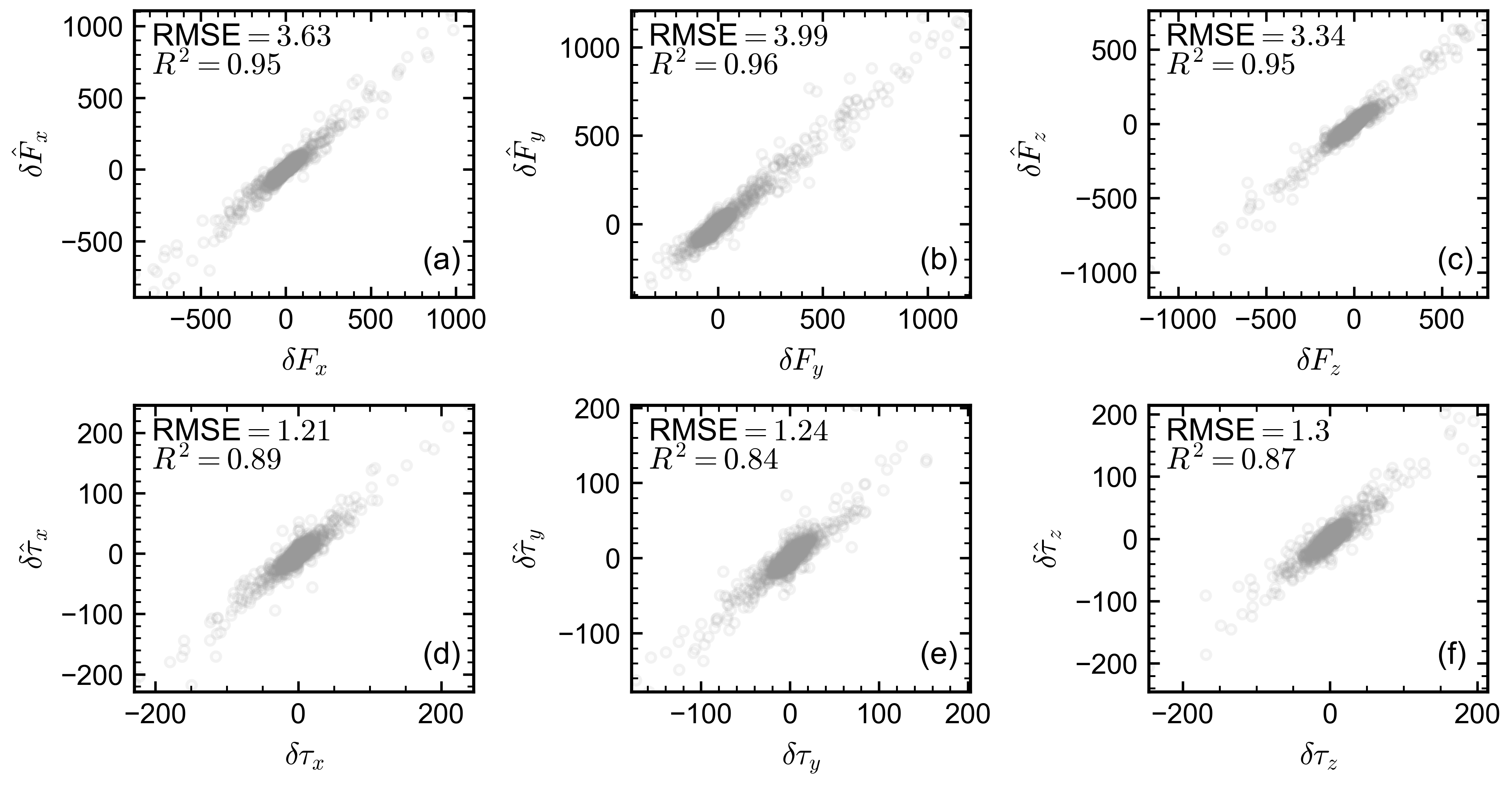}
        \caption{Same as Fig.~\ref{fig:ft_parity_benzo} but for phenoxide.}
\end{figure}

\begin{figure}[!h]
        \includegraphics{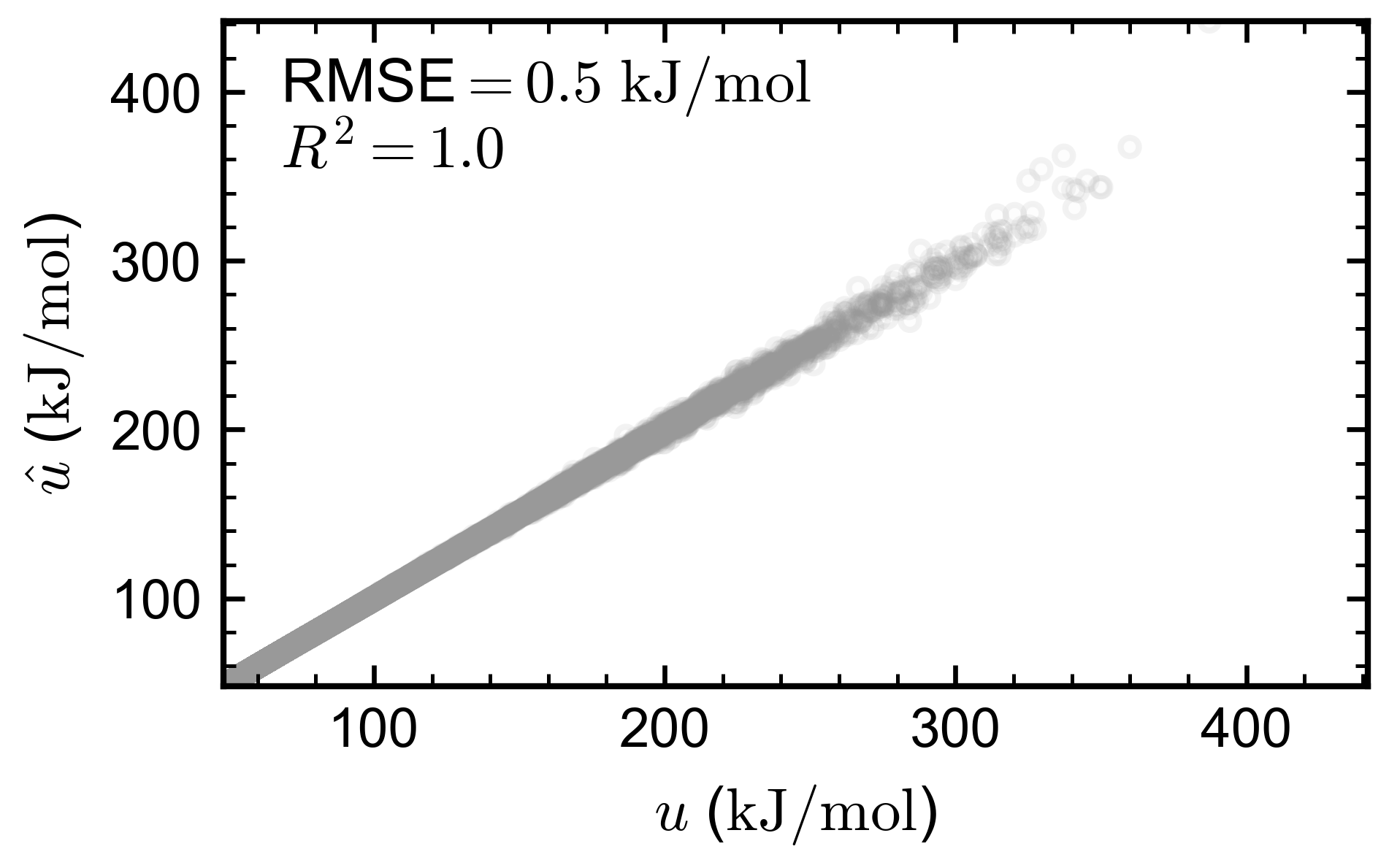}
        \caption{Same as Fig.~\ref{fig:ut_parity_benzo} but for phenoxide.}
\end{figure}

\begin{figure}[!h]
        \includegraphics[width=\linewidth]{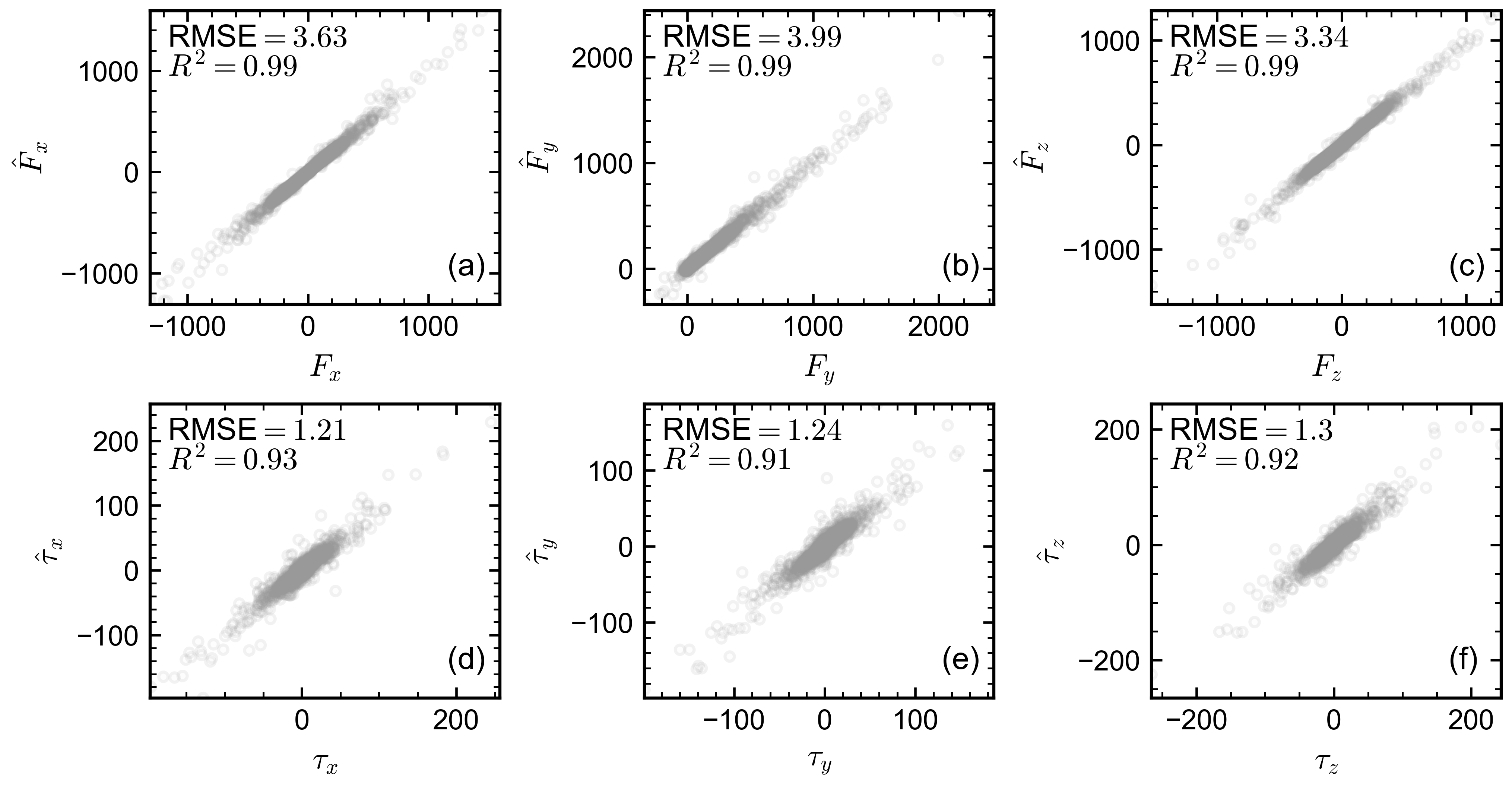}
        \caption{Same as Fig.~\ref{fig:ft_t_parity_benzo} but for but for phenoxide.}
\end{figure}